# Multicasting Optical Reconfigurable Switch

Niyazi Ulas Dinc[1, †], Mustafa Yildirim[1, †], Ilker Oguz[1], Christophe Moser[1], and Demetri Psaltis[2]

[1] Laboratory of Applied Photonics Devices, Ecole Polytechnique Fédérale de Lausanne (EPFL), Switzerland

[2] Ecole Polytechnique Fédérale de Lausanne (EPFL), Switzerland

† Equally contributing authors.

## Abstract

Artificial Intelligence (AI) demands large data flows within datacenters, heavily relying on multicasting data transfers. As AI models scale, the requirement for high-bandwidth and low-latency networking compounds. The common use of electrical packet switching faces limitations due to optical-electrical-optical conversion bottlenecks. Optical switches, while bandwidth-agnostic and low-latency, suffer from having only unicast or non-scalable multicasting capability. This paper introduces an optical switching technique addressing this challenge. Our approach enables arbitrarily programmable simultaneous unicast and multicast connectivity, eliminating the need for optical splitters that hinder scalability due to optical power loss. We use phase modulation in multiple layers, tailored to implement any multicast connectivity map. Phase modulation also enables wavelength selectivity on top of spatial selectivity, resulting in an optical switch that implements space-wavelength routing. We conducted simulations and experiments to validate our approach. Our results affirm the concept's feasibility, effectiveness, and scalability, as a multicasting switch by experimentally demonstrating 16 spatial ports using 2 wavelength channels. Numerically, 64 spatial ports with 4 wavelength channels each were simulated, with approximately constant efficiency (< 3 dB) as ports and wavelength channels scale.

Datacenters manage and process large volumes of data, comprising thousands of computing nodes alongside networking and storage systems. The networking fabric enables interconnection across many devices to execute workloads efficiently. Data routing within a datacenter relies primarily on electrical packet switching through fiber optic cables, where an electronic switch receives a data packet, processes, and regenerates it for the target port. In traditional network architectures, network devices themselves make decisions about where to forward data, based on pre-defined protocols. Meanwhile, network management has evolved, and software-defined networking (SDN) has been established to manage the datacenter network more efficiently, optimizing global traffic across the entire datacenter [1]. SDN separates the control plane (deciding where traffic should be sent) from the data plane (the actual forwarding of the traffic). Therefore, it is not necessary anymore to read packets, instead, the centralized controller can communicate the connectivity map to the switch. This evolution enabled the adoption of unicast (one-to-one) optical circuit switches (OCS) in datacenters [2-4].

Large AI models can occupy thousands of computer nodes, with model parameters and dataset are multicast among nodes iteratively during training [5, 6]. As the volume of multicast communication patterns increases along with the scale of operations, network congestion becomes unavoidable. Current optical switches cater solely to unicast, managing multicast via message replication with multiple electronic switches or passive power splitters coupled with programmable unicasting OCSs [4, 6-9]. However, the latter's power loss (due to multiple fixed divisions) curbs scalability, leaving a need for scalable and reconfigurable OCSs equipped with both unicast and multicast functionalities.

OCS based on microelectromechanical systems (MEMS) mirrors [10] is a deployed optical interconnect technology. While MEMS switches can provide energy-efficient optical routing, they cannot multicast a single input to many output channels. This arises from the ray-optic characteristics when using millimeter-sized, two-axis continuously rotating mirrors. Moreover, these systems are not able to switch among wavelength channels. [3, 4]. Another solution, Wavelength-selective switches (WSSs) are devices capable of redirecting optical signals to different output ports based on their wavelengths. Such switches typically consist of a one-dimensional (1D) diffraction grating and a beam-steering mechanism. The diffraction grating disperses the optical signals into their constituent wavelengths, while the beam-steering mechanism directs them to specific output ports based on their wavelengths. In many WSSs, Liquid-Crystal-on-Silicon (LCoS) spatial light modulators are utilized as beam-steering devices [11, 12]. However, the limitation of WSSs is that they can only redirect one input signal to one output port at a time. Consequently, WSSs are also unsuitable for multicasting. Furthermore, the 1D input/output configuration also limits the scalability of WSSs unless it is combined with MEMS-based re-routing at the cost of increased complexity [12]. Fixed interconnect methodologies, such as waveguides and volume holograms [14-18] are studied to address some of the mentioned challenges. However, they are non-programmable, restricting their applicability in dynamic environments like datacenters.

In this work, we introduce a multicasting optical reconfigurable switch (MORS) capable of multicast and wavelength selectivity. Our method leverages spatial light modulation to enable programming of optical paths consisting of both unicast and multicast connections (Fig. 1a). The current approach in industry for multicast employs fixed power splitters combined with optical unicast switches (Fig. 1b). An optical unicast switch, like MEMS-based mirror arrays, acts as a shutter to control light passage. In this approach, a switch having $N_i = N_o = N$ input/output ports splits each input light into N paths, regardless of the demanded multicast port count ($N_m$). Consequently, when the number of demanded multicast receivers is less than the output port count ($N_m<N$), power to the non-receiving $N-N_m$ output ports is inevitably wasted, slashing power efficiency. This limitation typically restricts such devices to a maximum of 16 output ports, which is significantly lower than the hundreds of ports that unicast switches can accommodate. [19].

There are two main methods for spatial light modulation: MEMS mirrors and wavefront shaping. We present these switching methods in a simplified manner, as shown in Fig. 1c and 1d. The primary difference lies in the fact that MEMS-based techniques directly connect the incoming beam to the output port via a single mirror, without altering the wavefront. Conversely, wavefront shaping enables more complex connections between input and output by utilizing multiple pixels. This allows for the realization of a multicasting switch using wavefront shaping, addressing a key power efficiency issue present in current multicast switches.

Multi-layer light modulation has been used for mode multiplexing by Morizur et al. using patterned glass [20, 21]. Multiple reflections off a liquid crystal Spatial Light Modulator (SLM) are used for volumetric computer-generated holograms [22], mode multiplexing [23-25], unscrambling output field after propagation through a multimode fiber [26], and optical neural networks [27]. In this study, we used an SLM with a mirror positioned across it to perform reconfigurable multicast switching. This arrangement permits the simultaneous  display of multiple modulation planes on a single SLM device, creating a multi-bounce single-pass cavity. Input beams traverse the multiple modulation planes/layers. In Fig. 1e, the spatial interconnect is illustrated, where different input ports are connected in unicast and multicast scenarios. Each 2D patch displayed on the SLM alters the phase of the light as it bounces and propagates through the layers. Fig. 1e displays a photograph of the experimental setup of MORS.

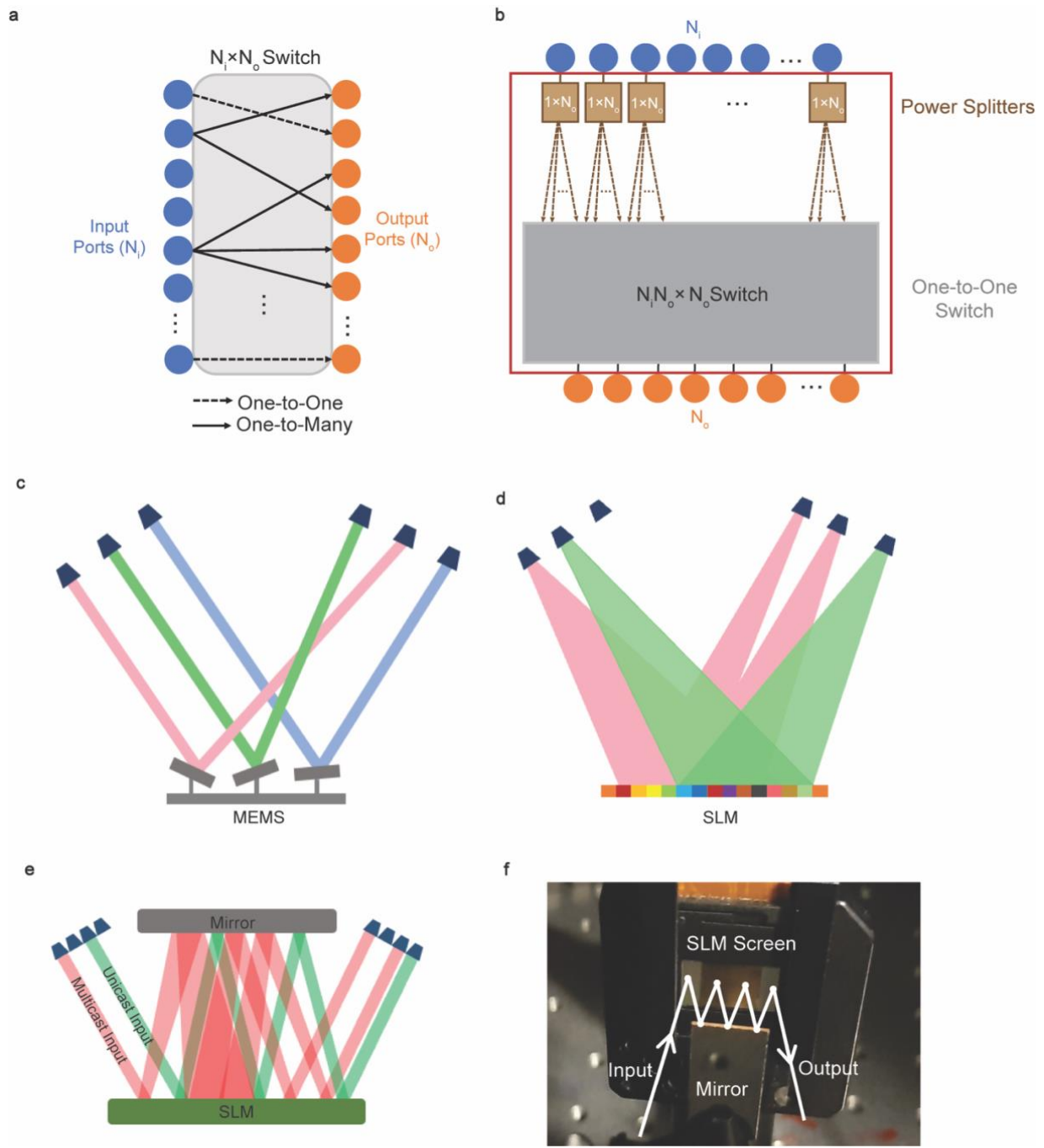

*Figure 1: Implementation of MORS.* a) Schematic representation of an optical switch capable of several concurrent unicast (One-to-One) and multicast (One-to-Many) connections. b) Current solution for obtaining a unicast and multicast capable switch by combining $N_o$ power splitters with a $N_iN_o \times N_o$ unicast optical switch. c) Schematic representation of a simplified mirror-based optical switch, which is only capable of one-to-one connections using a specific mirror per connection. d) Schematic representation of a simplified SLM-based optical switch, where connections are set collectively by the contributions of SLM pixels. Wavefront shaping enables multicast connections without power splitters. e) Conceptual implementation of MORS consisting of wavefront modulation over multiple planes where two incoming light beams are routed to different ports at the output plane with specific examples of unicast (One-to-One) and multicast (One-to-Two) connections. e) Laboratory realization of MORS using four modulation planes on a single SLM display.

# Results

The MORS technique is based on multiple evenly spaced phase-only modulation layers. We created a digital model of MORS using the Beam Propagation Method (BPM) to determine the phase patterns displayed on the SLM. We utilize an error backpropagation method thanks to having a differentiable forward model [28-30]. We refer this technique as Learning Tomography in our previous studies [14, 27, 28, 31], where more details can be found.

For clarity, we first explain the configuration and obtained results for multicast using a single wavelength channel . Later, in the section *Space-Wavelength Granularity,* we demonstrate  multiple wavelength channels. Finally, we investigate the scalability of this architecture by increasing the number of parameters when multicast and wavelength-selectivity is combined in the *Scaling study* section. In all the following sections, both experiments and simulations are the mean values of ten randomly selected connectivity maps to demonstrate a general trend. For further implementation details, see Methods.

1. ## Multicast

On-demand multicast is defined as splitting the data carrying light beam based on the number of destination ports and routing them accordingly (see Fig. 2a). We use the term "on-demand" to refer to the fact that the number of destinations and their positions can be changed arbitrarily. In Fig. 2b we show how the efficiency is severely hampered if one constructs a multicast switch based on ideal fixed splitters as the number of ports (N) scales ($N_i = N_o = N$). Such multicasting system is only 100% efficient when all the output ports receive data ($N_m = N$) as shown in Fig. 2e. On the other hand, a switch capable of performing on-demand multicast ideally should be lossless for any number of multicasts as shown in Fig. 2c, where the incoming power from each input port is split just by the number of required multicast output ports, $N_m$. MORS treats each connectivity map separately and generates the respective phase patterns. Therefore, input light is actively distributed with respect to the demanded $N_m$ rather than being set to N ports as in fixed splitters.

We conducted a numerical scaling study of MORS by varying the port count N from 4 to 128 for various multicast port count ($N_m$). The outcomes are presented in Fig. 2d. In comparison to a passive splitter, our observations reveal that MORS's power efficiency consistently surpasses that of the passive splitter when $N_m < N$. Notably, the efficiency declines as $N_m$ increases, except for the case where N equals $N_m$. In cases where N equals $N_m$, only one input port can be active at a time, whereas when $N_m$ is less than N, more than one input is active, resulting in the sharing of free parameters used to define the routing paths for different inputs. We found that a higher number of multicasting ports necessitates more parameters, resulting in reduced power efficiency for such ports. See the *Scaling Study* section for further analysis of this trend. We performed an optical experiment for N up to 16 and its efficiencies are presented in Fig. 2g. The experiments demonstrate a similar trend as in the simulations.

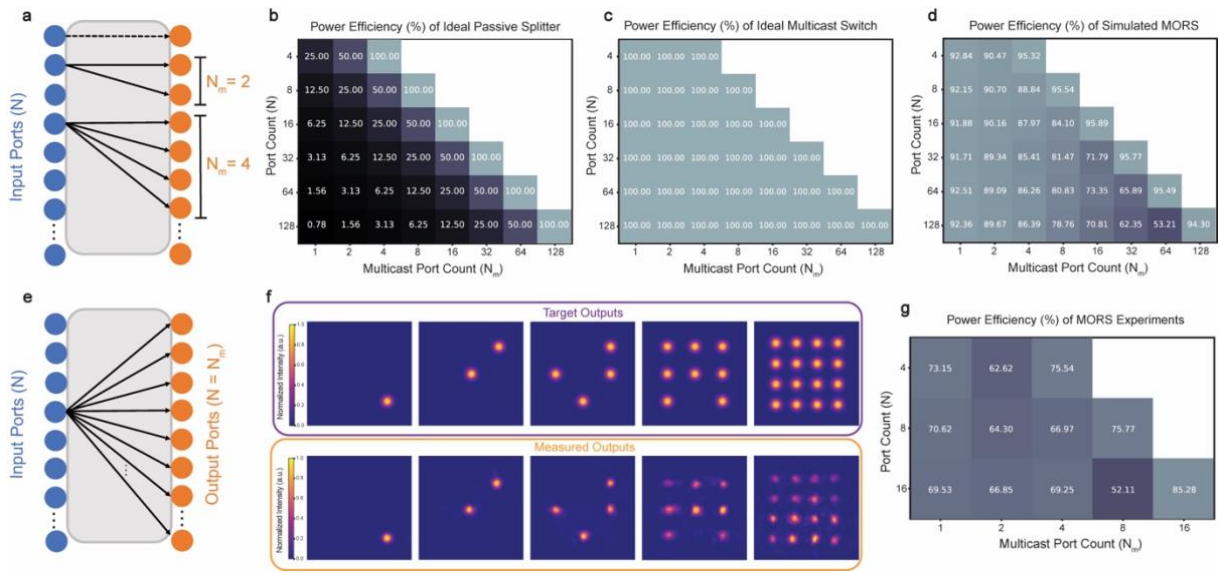

*Figure 2: description and comparison of fixed multicast and on-demand multicast along with numerical and experimental efficiency results of MORS.* a) Sketch of a switch that performs on-demand multicast and unicast where the number of multicast connections is less than the number of output ports b) Scaling of power efficiency of an ideal passive splitter with respect to the number of output ports (N) and utilized output ports ($N_m$) for multicast connections. c) On-demand multicast switch scales without loss ideally. d) The numerical results for scaling of MORS with finite degrees of freedom. e) Sketch of a switch where one input broadcasts its signal to all the output ports ($N_m$=N). f) Examples of experimental results obtained with MORS for on-demand multicast, given with corresponding target power distributions in the output plane. g) Experimental results for scaling of MORS that include the effects of finite degrees of freedom and experimental imperfections carried out up to N=16.

## 2. Space-Wavelength Selectivity

Wavelength division multiplexing is widely used in communication networks to enhance the capacity of spatial channels by aggregating signals with distinct wavelengths onto a spatial channel. WSS initially disperses an incoming beam carrying multiple wavelength channels and subsequently directs each channel to its target port. One can extend the spatial switch to accommodate the spectral switch by simply cascading both. Such an approach introduces more complexity and diminished energy efficiency. On the other hand, MORS can execute both space and wavelength simultaneously without any additional WSS layer by taking advantage of the wavelength dependence of light diffraction, meaning that wavelength channels entering from the same input port/space can be mapped to distinct spatial positions at the output based on their respective wavelengths as desired (see Fig. 3a). Moreover, MORS uses 2D input and output grids, which provides an additional dimension for scalability when compared to WSS, based on 1D input and output grids where the second dimension is reserved for dispersion. MORS is a 3D device thanks to the utilization of multiple 2D modulation planes, which permits the use of 2D input and output grids.

There are two possible switching scenarios: either all wavelengths have the same connectivity map (i.e. broadband response), or the connectivity of all wavelength channels is arbitrarily chosen (i.e. wavelength-selective response). We can determine the phase masks that implement wavelength-selective connections by performing simulations where we run the forward model for each wavelength. After the optimization step, we can also probe the wavelength response numerically. We start by training using a single wavelength (850 nm) for a switch that has N=4 input and N=4 output ports. We summarize the performance using an efficiency-crosstalk plot. We place the correctly routed powers on the diagonal while the off-diagonal values show the crosstalk (Fig. 3b). Numerical results show –0.22 dB mean insertion loss and –30.71 dB mean crosstalk. In Fig. 3c, we demonstrate the wavelength response of the device, which is relatively broadband, spanning roughly 200 nm range with efficiency staying above 50%. If we use five wavelengths during training as indicated in the plot in Fig. 3c, and set the connectivity maps the same for all the wavelengths, we see that we can achieve even more flat response with a few percent efficiency drop in the central wavelength as a trade-off.

For the wavelength-sensitive switch, we assign four wavelength channels that are 30 nm apart where each wavelength has a different, arbitrarily chosen connectivity map. In this case, we have four efficiency-crosstalk plots for each wavelength channel as shown in Fig. 3d. We plot them on a bigger matrix, which gives an overview of allowed conversions among spatial ports and wavelength channels. The hatched off-diagonal parts are not allowed since in a linear system there is no mechanism for frequency conversion. The mean insertion loss is -0.59 dB, and the mean crosstalk is -20.22 dB. In Fig. 3e, we provide the wavelength response for a specific connectivity map where the spatial output ports do not overlap for ease of representation of the individual wavelength channels. From Fig. 3e we see that there is a huge bandwidth allocation to individual wavelength channels considering that 1 nm corresponds to >400 GHz at 850 nm central wavelength. For the experiments, we show the wavelength selectivity for two wavelengths at 835 nm and 865 nm and provide the experimentally obtained efficiency-crosstalk plot in Fig. 3f. In these experiments, we obtained -2.10 dB insertion loss and -19.12 dB crosstalk.

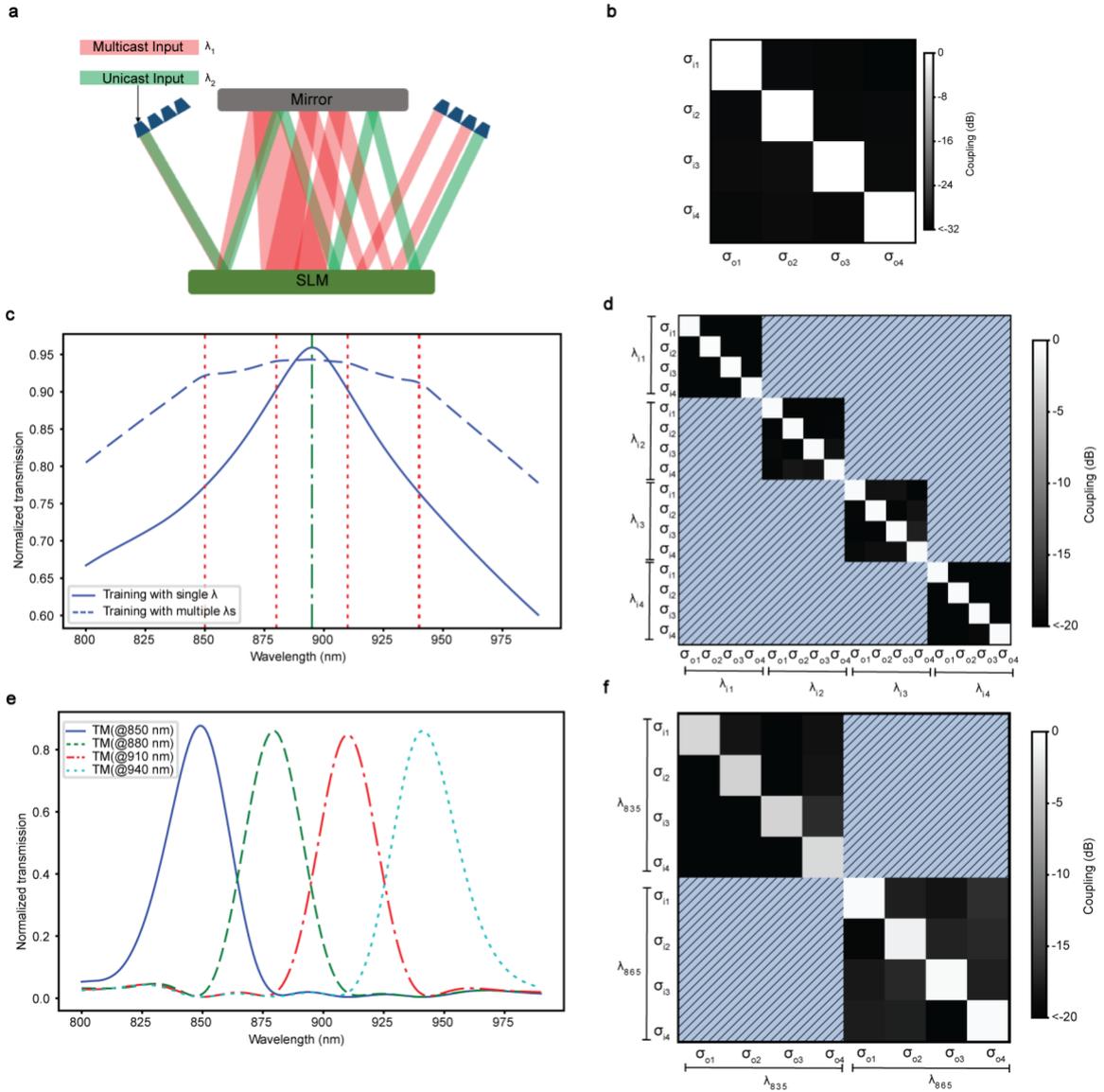

*Figure 3: Efficiency and crosstalk results of MORS for broadband and wavelength-selective operation.* a) Schematic representation of MORS showcasing wavelength selective unicast and multicast connections. b) The efficiency-crosstalk plots (b, d, f) show the correctly routed powers on the diagonal from the $n^{th}$ spatial input port $\sigma_{in}$ to the corresponding $n^{th}$ spatial output port $\sigma_{on}$ for each wavelength. Off-diagonal elements are crosstalk between the target and other ports. Simulated efficiency-crosstalk plot for a single wavelength with four spatial ports is depicted, providing insights into the connectivity map. c) Average optical power transmission for broadband application (covering the same connectivity over a 100nm range) is illustrated, we present power transmissions for both single and multiple wavelength training cases. d) The simulated efficiency-crosstalk plots for four different wavelengths are presented, showing distinct connectivity maps for each wavelength channel, all sharing spatial ports. e) Power transmission of four different wavelengths is shown while each wavelength has a different connectivity map. f) Experimentally measured efficiency-crosstalk plots for two wavelength channels (835nm, 865nm), demonstrating the practical application of MORS.

## 3. Scaling study

For the implementation of large networks, the scalability of MORS is of paramount importance. Therefore, we conducted a scaling study based on simulations by devising a switch that has 64 spatial input ports and 64 spatial output ports operating with four different wavelength channels. This results in 256 spatiotemporal input and 256 spatiotemporal output ports. We calculate the parameter count of MORS by multiplying the layer number by the number of pixels used in one layer, which results in the total number of pixels (see Methods). In Fig. 4a, we plot the mean efficiency results for different configurations. We clearly see that the increased degrees of freedom improve the overall performance. We observe that while the 2-layer configuration performs poorly, 6-layer and 8-layer configurations settle in a comparable scaling trend, showing a direct link between employed parameters and achievable efficiency once the system has sufficient depth to use diffraction to separate and route different wavelengths to their corresponding spatial output ports. In Fig. 4b, we present signal to noise ratio (SNR) which is coupled optical power over crosstalk in dB scale. We observe that although the coupling efficiency levels off around three million parameters/pixels, SNR continues to improve with the additional parameters.

Next, we investigate how parameters/pixels are shared among the different connections, where a connection is a link from a specific input to one or more desired output ports. We define allocated pixels by counting how many pixels are used when input light is routed to its target port(s). In Fig. 4c, we illustrate an example of allocated pixels corresponding to the multicasting configuration up to 64 (the last row of Fig. 2d). The pixel values of Fig. 4c indicate how many connections go through each pixel. In Fig. 4d, we show how well each connection works based on how many pixels it uses for each output port. For example, a connection multicasting to 64 ports uses around 400,000 pixels, which results in roughly 6,000 pixels per output port as shown in Fig. 4d. The data reveals an inverse relationship between the connection efficiency and its multicast port count. While connections with higher multicasting use more pixels overall, the number of allocated pixels per output port is lower.

Furthermore, significant variations arise between different random mappings for large multicast counts (i.e. 32 or 64) indicated by the same color-code in Fig. 4d. This suggests that efficiency is impacted not only by the number of allocated pixels but also by their overlap. We define overlap as the degree to which parameters/pixels allocated by a specific connection are shared with other connections (see Methods). In Fig. 4e, we plot efficiency against the overlap calculated from the data points of Fig. 4d. The results demonstrate that higher overlap (more shared pixels) leads to a lower connection efficiency. This trend becomes increasingly pronounced for connections with larger multicast port counts. For example, the yellow data point with the efficiency of >70% shown in Fig. 4d is located on the left-most side of Fig. 4c, indicating a small amount of overlap for this particular connection.

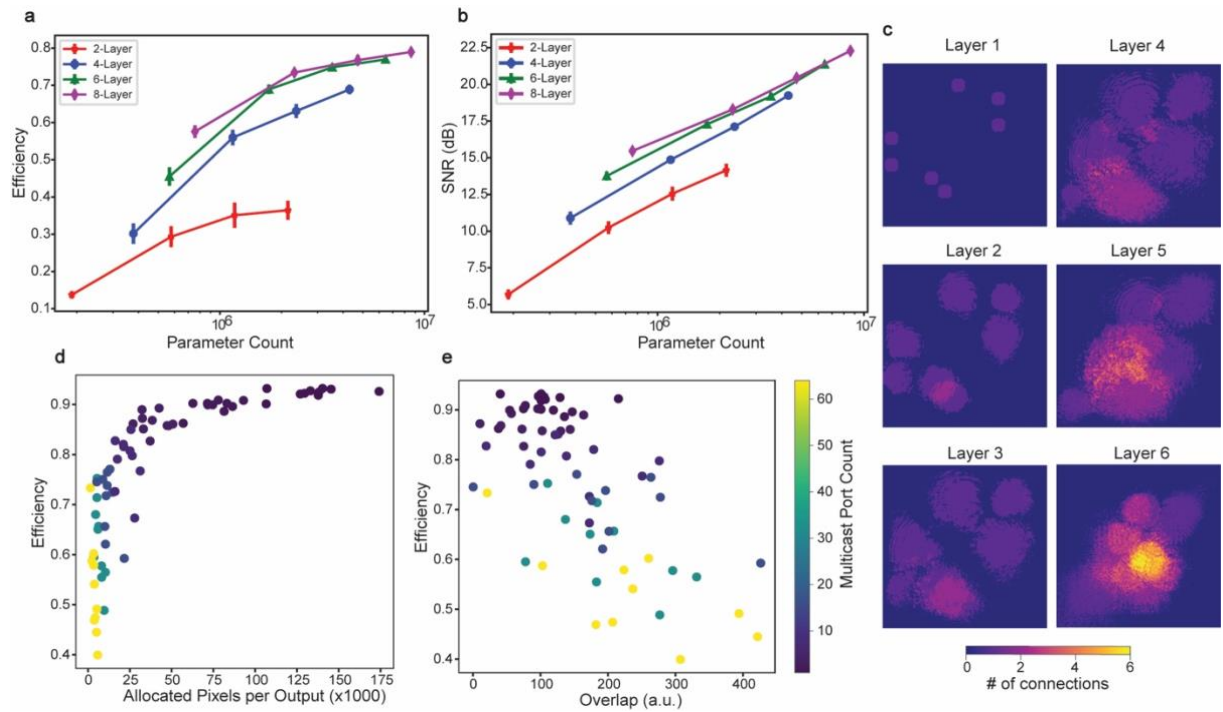

*Figure 4: Efficiency trend of MORS with respect to the number of pixels (parameter count). a) Average coupling efficiencies for various number of layers are plotted while parameter count is varied by changing layer width. b) Scaling of SNR for the same configuration as in a. c) An example of allocated pixels over six layers. The color code measures how many connections pass through each pixel. d) The efficiency of individual connections for a single wavelength switch having 128 ports. The number of multicasting outputs is up to 64. The data for ten different (randomly picked) connectivity are present. X-axis denotes the number of pixels that are actively used by these connections. c) The efficiency versus overlap (shared pixels) plot for the same device as in d. Efficiency follows a decreasing trend with higher overlap. In d and e, color code represents multicast port count.*

## Discussion

MORS is different from conventional methods by introducing a novel optical path reconfiguration employing spatial light modulation across multiple planes. This approach inherently incorporates multicasting capabilities with wavelength selectivity. The promising scaling trends we obtained from numerical studies were substantiated by experimental validation. It is noted that free-space techniques inherently lead to efficient throughput, a key advantage over integrated solutions where factors like waveguide loss and coupling losses can impact scaling. Moreover, the ability to operate on the wavefront facilitates optimization with respect to the étendue of the output fibers. For simplicity, this aspect is ignored in the drawings shown in Fig. 1c and 1d. MEMS-based systems require two separate mirror arrays, the first one to define the output position and the second one to correct the beam angle to couple light. As the beam size gets larger along with the device scale, respecting the étendue becomes challenging without precise angle control of the mirrors [4]. This is one of the reasons limiting the scalability of MEMS switches [19]. MORS on the other hand has more degrees of freedom to shape the wavefront, optimizing étendue of the beam for coupling.

Our scaling analysis demonstrated a consistent improvement in the SNR with an increasing number of pixels. Notably, at around a million pixels, deeper architectures lead to enhanced performance. However, our findings also revealed a performance bottleneck stemming from shared pixel usage among connections, highlighting the limitations of solely increasing parameter count for scaling.

With MORS, the efficiency of multicasting remains consistent, offering a potential resolution to longstanding challenges for upscaling multicasting optical switching. Our approach also has the capability to manage both space and wavelength. To conclude, MORS offers a new perspective in network design, potentially bridging the efficiency gap in multicasting present in today's optical interconnect technologies with additional flexibility by possessing space-wavelength granularity.

## Methods

**Experimental setup:** For our experiments, a continuous wave Solstis M2 laser was employed. The selected mirror, measuring 11.6 mm in width and 17.1 mm away from the SLM display, accommodates the four reflections. To channel the input beam to the SLM, 4F imaging was applied, relaying the beam that was reflected off a digital micromirror device (DMD). This DMD acts to simulate the input grid by activating specific sub-regions. For the modulation layers, we designated patches of 280 by 280 pixels on the SLM. The SLM used in our arrangement has a pixel pitch of $\Lambda=9.2$ μm.

Following the fourth reflection, the beam is imaged, with the resultant output intensity captured by a CMOS camera. For the wavelength-selective experiments, the tunable laser is utilized to test the results at two wavelengths that are 30 nm apart (835 nm and 865 nm).

**Multicast and space-wavelength granularity:** In numerical studies showing multicast in the first part of the *Results* section, we employed six diffractive layers. The layer width is set to 640 pixels for the multicast case. To achieve wavelength selectivity, we execute the forward model for each desired central wavelength, optimizing the pixel phase based on the aggregated backpropagated error signals. We refer to these wavelengths as training wavelengths. For wavelength-selective numerical experiments, we also used six diffractive layers having a layer width of 360 pixels. Layer-to-layer distance is set to four millimeters for both cases. The details about the training method can be found in [27].

For the scaling study, we used the layer width values 307, 538, 768, and 1036. The number of layers is swept from 2 to 8. We activated 20 input ports for each wavelength channel randomly. By exceeding 64 available spatial ports, we ensured that some of the ports transmit multiple wavelengths. For the corresponding outputs, we populated all the possibilities, meaning that there are 256 active outputs distributed among 64 spatial ports and 4 wavelength channels. Each input is mapped randomly to the output ports including unicast and multicast connections, where the number of multicasts for each input is also arranged randomly. In other words, a single input is mapped to output ports where the output port number is in the range of 1 to 44. These selections are restricted to the available number of output ports and channels.

**Overlap (shared pixels)**: We calculate the overlap amount to indicate shared pixels by tracing each input over the modulation planes and label pixels where the light intensity is higher than a certain threshold as a "allocated pixel" for that input. We iterate this procedure for all the inputs and obtain allocated pixels for every connection. Next, we assign a degree to pixels according to the number of connections that it contributes to. For a pixel that only contributes to a single connection, this degree is zero. The overlap is defined as sum of the assigned degrees of allocated pixels, divided by thousand.